\documentclass[11pt]{article}
\usepackage{amsmath, amssymb, amsthm, float, fullpage, graphicx, multirow,  parskip,  subcaption}
\usepackage{tikz}

\usepackage{hyperref}
\usepackage{natbib}

\definecolor{PennRed}{RGB}{152, 30 50}
\definecolor{PennBlue}{RGB}{0, 44, 119}
\definecolor{PennGreen}{RGB}{94, 179,70}
\definecolor{PennViolet}{RGB}{141, 76, 145}
\definecolor{PennSkyBlue}{RGB}{14, 118, 188}
\definecolor{PennOrange}{RGB}{243, 117, 58}
\definecolor{PennBrightRed}{RGB}{223,82, 78}

\hypersetup{pdfborder = {0 0 0.5 [3 3]}, colorlinks = true, linkcolor = PennBrightRed, citecolor = PennSkyBlue}

\title{Protocol for an Observational Study on the Effects of Playing High School Football on Later Life Cognitive Functioning and Mental Health}
\author{Sameer K. Deshpande\thanks{The Wharton School, University of Pennsylvania. These authors contributed equally.}, Raiden B.Hasegawa\footnotemark[1], Amanda R. Rabinowitz\thanks{Moss Rehabilitation Research Institute}, John Whyte\footnotemark[2],\\ Carol L. Roan\thanks{University of Wisconsin -- Madison}, Andrew Tabatabaei, Michael Baiocchi\thanks{Stanford University}, \\ Jason H. Karlawish\thanks{Perelman School of Medicine, University of Pennsylvania}, Christina L. Master\thanks{Children's Hospital of Philadelphia}, Dylan S. Small\thanks{The Wharton School, University of Pennsylvania. Corresponding author. Email: \nolinkurl{dsmall@wharton.upenn.edu}}}
\date{7 July 2016}

\begin{document}
\maketitle

\begin{abstract}
\noindent \textbf{Background}: Chronic Trauma Encephalopathy (CTE) is a neurodegenerative disease that is the result of repetitive brain trauma. 
Symptoms may include memory loss, aggression, confusion and depression. 
In recent years, CTE has been the focus of both investigative reporting and research examining the neurological consequences of cumulative head trauma to which many professional football players are exposed. 
A potential causal relationship between head injuries suffered by NFL players and later-life neurological decline may have far-reaching public health implications for participants in youth and high school football programs. 
American football is the largest participation sport in US high schools and is the leading cause of concussion among adolescents and young adults. 
However, brain trauma risk at the professional level may be different than that at the youth and high school level and the long-term effects of participation at these lower levels is as-yet unclear.
To investigate the effect of playing high school football on later life depression and cognitive functioning, we propose a retrospective study using data from the Wisconsin Longitudinal Study (WLS) of graduates from Wisconsin high schools in 1957.

\noindent \textbf{Methods and Analysis}: Our proposed study is a retrospective observational study that compares 1,153 high school males who played varsity football with 2,751 male students who did not. 1,951 of the control subjects did not play any sport and the remaining 800 controls played a non-contact sport. 
We focus on two primary outcomes measured at age 65: a composite cognitive outcome measuring verbal fluency and memory and the modified CES-D depression score. 
To control for potential confounders we adjust for pre-exposure covariates such as IQ with matching and model-based covariate adjustment. 
We will conduct an ordered testing procedure designed to use the full pool of 2,751 controls while also controlling for possible unmeasured differences between students who played sports and those who did not.
We will quantitatively assess the sensitivity of the results to potential unmeasured confounding. 
The study will also consider several secondary outcomes of clinical interest such as aggression and heavy drinking.

\noindent \textbf{Discussion}: This retrospective observational study will contribute to the current public health research interest in youth and high school football participation with evidence of whether playing high school football has a causal effect on later-life neurological outcomes. The rich set of pre-exposure covariates, relatively unbiased sampling, and the long-term longitudinal nature of the WLS dataset make the proposed analysis unique among related studies that primarily rely on convenience samples of football players who have reported neurological symptoms. 


\end{abstract}

\section{Background and Motivation for Study}
\label{sec:background_motivation}

American football is the largest participation sport in U.S. high schools Ð more than 1 million boys played high school football in 2014. 
Playing football involves many hits to the head and is associated with a high rate of concussion of the brain. 
Football is a leading cause of concussion among adolescents and young adults \citep{Guskiewicz2000, Gessel2007}, accounting for nearly half of all sports-related concussions \citep{Marar2012}. 
These estimates may, in fact, underestimate the true prevalence of football-related head injury, as evidence suggests that as many of a half of all injuries may go unreported \citep{McCrea2004}
Professional football players are 3 times more likely to die of a neurodegenerative disease than the general U.S. population \citep{Lehman2012}.
Among professional players who have donated their brain for study at death, 87 of 91 have been found to have chronic trauma encephalopathy (CTE), a neurodegenerative disease resulting from repetitive brain trauma whose symptoms may include memory loss, aggression, confusion and depression \citep{Breslow2015}.
Although professional players donating their brain for study at death is a biased sample, the high rates of CTE in this specialized sample have raised concerns about the cumulative effect of blows to the head inherent in football on later-life cognition and mental health. According to the American Association of Pediatrics statement on tackling in youth football, Òthe recognition of these injuries and the potential long-term sequelae have led some physicians to call for a reduction in the number of contact practices, a postponement of tacking until a certain age, and even a ban on high school football.Ó 

While it is possible that playing high school football increases the risk of later-life cognitive impairment and depression, most high school football players ultimately have less years of exposure to football-related head trauma. 
As such, it is possible that high school players are not at as high of a risk as professional players. 
Moreover, playing high school football can positively affect personal development, through its emphasis on instilling leadership, responsibility, perseverance, and teamwork, as well as conferring the benefits of regular exercise on overall health. 
These positive effects could counter-balance the negative effects associated to repeated head trauma. 
Ultimately, there is little compelling evidence either way of the potential benefits or harms of playing high school football on later life mental health and cognition, motivating the present study.

In theory, the Ògold standardÓ approach for establishing a causal relationship between playing high school football and later-life depression and cognitive impairment, would be to randomly divide high-school students to two groups, one assigned to play football and one assigned to not play football, and to compare the later life mental and emotional health of the students in each group. Such a study, of course, is highly impracticable and we instead must rely on observational data. The main challenge in estimating the effect of a treatment (in this case playing high school football \footnote{More precisely, playing high school as it was played in Wisconsin in the mid-1950's.}) on a given outcome (later life mental health and cognitive impairment) is that the treated and control groups are generally not comparable prior to treatment. As a result, any observed difference in outcomes between the two groups may or may not be attributable to the treatment itself. A common statistical solution to this problem is to first match each treated subject to control subjects along several baseline covariates and then to compare the outcomes within each matched set.

A standard criticism of such an approach is that while we may control for several observed potential confounders, there are still myriad unobserved confounders for which we have not controlled. This prompts us to conduct a sensitivity analysis. In short, a sensitivity analysis examines departures from random treatment assignment that a potential confounder may introduce and the effects such departures have on inferences about treatment effects.

We propose to use data from the Wisconsin Longitudinal Study (WLS) of graduates from Wisconsin high schools in 1957 \citep{HerdCarrRoan2014} to investigate the link between playing high-school football and later life depression and cognitive impairment. 
The WLS has a number of attractive features that make it well-suited for such a study. 
First, it records whether study participants participated in high school football and also includes detailed measurements of later-life mental health, psychological well-being, and cognition. 
Second, it includes a rich set of baseline covariates which we may use to construct matched sets of treated and control individuals, including family background, adolescent characteristics, educational and occupational achievement and aspirations. 
Third, the WLS is one of the few longitudinal data sets that includes an administrative measure of childhood cognition. 
In short, the WLS provides a large data set that facilitates comparing the later life mental health and cognitive ability of men who played high school football to those who did not, after carefully controlling for a range of potential confounders.


\section{Eligibility and Exclusion Criteria}
\label{sec:eligibility_exclusion_criteria}

There are a total of 10,317 subsets in the WLS dataset. 
To determine whether a subject played football, we use data recorded from their senior year high school yearbook. 
In all, we are missing yearbook information \footnote{Either no yearbook was available for that school, the yearbook did not contain any student activity information, or the subject as not included in his or her yearbook} for 1,205 subjects (11.68\%), who are dropped from our analysis. 
Additionally, we exclude the 843 students for whom yearbook information was available but whose activity participant was not recorded under their senior photo or an in index\footnote{These students came from so-called ``complex schools'' for which activity information was not listed under senior photos or as part of an index. Instead, WLS coders had to rely on group pictures of teams and clubs to impute participation data. We find that there is no significant difference in the size of complex and non-complex schools but that the rate of football playing in complex schools is half that in non-complex schools. We suspect that there is misclassification of football playing in complex schools and we therefore drop students from these schools from our analysis.}. 
Of the remaining 8,296 subjects, we consider only the 3,973 (48.04\%) males. 
Since it is possible to suffer mild traumatic brain injuries or repeated concussive impacts in sports such as soccer, hockey, lacrosse, and wrestling, we exclude those subjects who did not play football but played on these ``risky'' sports\footnote{Out of the 2,820 subjects who did not play football, none played lacrosse or soccer, 6 played hockey, and 63 wrestled. We exclude these 69 subjects.}. 
After this exclusion, we are left with a total of 3,904 subjects, of whom 1,153 (29.53\%) played football (our treatment group) and 2,751 who did not play football or any other risky sports (our control group).
1,951 control subjects did not play any high school sport (in the sequel, we refer to these subjects as the Non-Sport Control group) and the remaining 800 subjects played a non-contact sport (we refer to these subjects as the Other Sport Control).

\section{Study Outcomes}
\label{sec:primary_outcomes}

We will consider two primary outcomes, one related to cognitive functioning and one related to depression, and several secondary outcomes. 
The WLS administered a battery of cognitive tests and mental health surveys in 1993, between 2003 and 2005, and between 2011 and 2013, when the subjects were approximately 54, 65, and 72 years old, respectively. 
Table~\ref{tab:outcome_availability} shows the available cognitive and depression outcomes and the number of eligible subjects for whom we have recorded each outcome. We see that the 2003-05 collection wave was the first to report results from the majority of cognition tests. 
Thus, for our primarily analysis, we focus only on test results from this collection wave.

\begin{table}[H]
\centering
\caption{Availability of cognitive test and depression scores in each collection wave. Percentage of eligible subjects for whom the outcome is available is shown in parentheses}
\label{tab:outcome_availability}

\begin{tabular}{lccc}\hline

Test & 1993 Wave & 2003-05 Wave & 2011-13 Wave \\ \hline
CES-D \footnote{Modified CES-D psychological distress/depression score} & 2747 (63.90\%) & 2643 (61.48\%) & 2005 (46.64\%) \\
Letter fluency (LF) \footnote{Subjects were asked to name as many words beginning with either ``L'' or ``F'', where ``L'' or ``F'' was randomly selected.} & 0 (0.00\%) & 1942 (45.17\%) & 2130 (49.55\%) \\
Delayed Word Recall (DWR) \footnote{Subjects asked to recall words nine minutes after ten words were read to them.} & 0 (0.00\%) & 2057 (47.85\%) & 1819 (42.31\%) \\
Similarities (SIM)\footnote{Subjects asked to say in what ways two objects are alike (e.g. ``In what ways are an orange and banana alike?'').} & 3407 (79.25\%) & 2727 (63.43\%) & 2337 (54.36\%) \\
Digit Ordering (DO) \footnote{Subjects asked to mentally rearrange and verbally restate increasingly lengthy sets of one-digits numbers in increasing order} & 0 (0.00\%) & 2097 (48.78\%) & 1592 (37.03\%) \\
Number Series (NUM) \footnote{The McArdle and Woodcock number series task measures indication and reasoning ability with particular emphasis on quantitative reasoning. Each item consisted of a series of numbers (e.g. 23,26, 30, 35, --) and subjects were asked to identify the number that correctly completed the series.} & 0 (0.00\%) & 0 (0.00\%) & 2332 (54.25\%) \\
Immediate Word Recall \footnote{Subjects were asked to remember and orally repeat as many words as they can from a set of ten. The immediate recall is asked right after the questioner reads the words. The same words are used in the DWR test.} & 0 (0.00\%) & 2135 (49.66\%) & 1820 (42.34\%) \\
Category Fluency (CF)\footnote{Subjects were asked to name as many items as they can think of in one of the randomly selected categories, ``Foods'' or ``Animals.''} & 0 (0.00\%) & 1281 (29.80\%) & 1095 (25.47\%) \\ \hline
\end{tabular}
\end{table}

\citet{Yonker2007} studied the the factor structure of the cognitive functioning test results in the WLS when respondents are 65, and identified three factors -- memory/attention (word recall and digit ordering), abstract reasoning (similarities) and verbal fluency (letter fluency and category fluency). 
For our primary outcome, we focus on one measure of memory/attention, delayed word recall and one measure of verbal fluency, letter fluency. 
These cognitive domains are most consistent with the National Institute of Health (NIH) recommendations for Common Data Elements in Traumatic Brain Injury research \citep{Wilde2010}, and have demonstrated the greatest sensitivity to sports-related concussion \citep{BelangerVanderploeg2005}. 
For our primary depression outcome, we use the modified CES-D score.
Table~\ref{tab:primary_availability} shows the availability of letter fluency score, delayed word recall score, and modified CES-D in our treated group and our two control groups. 
If football playing status affect the availability of test results (e.g. playing football increased the likelihood of dying young or early onset of debilitating cognitive impairment so that the subject was unable to participate in the WLS surveys), any comparison of treated and control groups based on these results will be biased.
To examine whether football playing status affected the availability of the test results, we fit separate logistic regression models to predict the availability of the letter fluency score, delayed word recall score, and modified CES-D score at age 65 using all of our baseline covariates along with football playing status. 
We find that football playing status was not a significant predictor of whether a subject will drop out of the study prior to observing later life cognitive or depressive outcomes.

\begin{table}[H]
\centering
\caption{Availability of Primary Outcome Components}
\label{tab:primary_availability}

\begin{tabular}{lcccc} \hline
Available Scores & Football & Non-Sport Controls & Other Sport Controls & All Controls \\ \hline
LF, DWR, CES-D & 467 & 682 & 301 & 983 \\ 
LF, DWR & 58 & 118 & 40 & 158 \\ 
LF, CES-D & 24 & 37 & 19 & 56 \\
DWR, CES-D & 55 & 92 & 33 & 125 \\ 
LF & 9 & 17 & 4 & 21 \\
DWR & 13 & 14 & 10 & 24 \\
CES-D & 210 & 332 & 159 & 491 \\
None & 319 & 659 & 234 & 893 \\ \hline
\end{tabular}
\end{table}

We construct our primary cognitive outcome by averaging the z-scores for LF and DWR. 
Of course, as suggested by Table~\ref{tab:primary_availability}, there are several subjects for whom one of the two primary cognitive tests scores is unavailable. 
For these subjects, we define their primary cognitive outcome to be the available z-score. 
We construct our primary depression outcome as the z-score for CES-D. 

In addition to these primary outcomes, we will also study several long-term secondary outcomes:
\begin{itemize}
\item{Primary cognitive and depression outcomes measured at age 72}
\item{Each cognitive test score and CES-D score at all available ages \footnote{We will not analyze CF because it was only administered to 50\% of the subjects who were subjects to the other cognitive tests}}
\item{Hostility Index at all available ages}
\item{Speilberger Anxiety Index at all available ages}
\item{Speilberger Anger Index at all available ages}
\item{Heavy drinking status at ages 54, 65, and 72 \footnote{A subject is classified as a heavy drinker if he reports having had more than 5 drinks on more than 5 separate occasions in the month preceding the interview.}}
\end{itemize}

We will also look at intermediate outcomes that might or might not be expected to be affected by playing football:
\begin{itemize}
\item{Duncan SEI score of occupational prestige for job held in 1964, 1970, and 1975\footnote{The score for 1964 is reported on the 1950-basis Duncan SEI scale while all other scores are reported on the 1970-basis scale.}}
\item{Subject's total earnings in 1974}
\item{Binary indicator for regularly engaging in vigorous physical activity at age 35}
\end{itemize}

\section{Matching}
\label{sec:matching}
We will compare the primary outcomes of the treated subjects to the primary outcomes of the control subjects, after controlling for baseline covariates via full matching with a propensity score caliper. 
Appendix~\ref{sec:baseline_covariates} lists the baseline covariates on which we match.
The output of the matching procedure is a collection of matched sets consisting of either one or more treated units and a single control unit or a single treated unit and one or more control units. 
This type of full matching is the optimal sub-classification for an observational study \citep{Rosenbaum1991}.

An intuitive first attempt at matching would be to group treated subjects with those controls subjects whose baseline covariate are identical. When there are several baseline covariates, however, such a strategy is difficult if not impossible to implement. Instead, we aim to create matched sets in such a way that for each covariate, the mean value of the covariate among the matched treated subjects is similar to the mean value of the covariate among the matched control subjects. This is known as covariate balance and we can assess the suitability of the match by looking at the standardized difference in the mean of each covariate between treated and control groups.  Minimizing a robust Mahalanobis distance with a propensity score caliper matching is one way to achieve adequate covariate balance \citep[see][Chapter 8]{Rosenbaum2010}.

As we saw in Table~\ref{tab:primary_availability}, not all individuals have all three primary test results available. 
To facilitate appropriate comparisons between treated and control groups, we first stratify our study population based on the availability of the primary outcome components (i.e. the rows of Table~\ref{tab:primary_availability}) and construct a match within each stratum. 
This ensures that all subjects in each matched set have the same test scores available. 

We hypothesize that, after adjusting for measured baseline covariates, the effects of playing football are long term and will not affect outcomes shortly after high school. If we saw differences among football and non-football players in outcomes shortly after high school after accounting for measured baseline covariates, we would be concerned that this may reflect unmeasured baseline differences between football and non-football players rather than an effect of football.  
Specifically, we consider the outcomes of years of education and military service.  
Since additional years of higher education can have protective effects on later-life cognition and experiences in the military may affect psychological well-being, differences in post-secondary education and military service between the football and non-football subjects would make it difficult to attributed any differences in primary outcomes to participation in high school football since we hypothesize that football does not affect post-secondary education after controlling for measured covariates.  
We perform Mantzel-Haenszel tests on binary indicators of military service up to 1975 and realized years of post-secondary education\footnote{In 1975, the WLS recorded each subject's equivalent years of education (12 for those who did not attend college, 13 for those with one year of college, etc.). Those with baccalaureate degrees were coded as having the equivalent of 16 years of education, those with 2-years masters degrees were coded as having 18 equivalent years, etc.}. 
We find no significant association between playing football and serving in the military or the amount of post-secondary education completed, after adjusting for the measured covariates. 
Thus, these tests provide no evidence of unmeasured baseline differences football and non-football players. 

\section{Testing in Order for Primary Outcomes}
\label{sec:testing_in_order}

Once we match treated subjects to control subjects, we can compare the two primary outcomes from the treated and control groups. 
However, there may still be residual covariate imbalances within matched sets. 
Although full matching can help eliminate bias by balancing covariates on average, some bias may remain due to these residual covariate imbalances. 
One way to further reduce bias as well as variance due to residual covariate imbalances is to combine full matching with covariate adjustment via regression \citep{Rosenbaum2002a, Hansen2004}.
In particular, we regress each primary outcome on matched set indicators, covariates, and a treatment indicator. 
In general, combining matching and model-based covariate adjustment methods have been shown to produce treatment effect estimates with lower bias than either method alone \citep{Rubin1973, Rubin1979}. 
We then test the hypothesis of no treatment effect using standard OLS p-values for the coefficient on the treatment indicator. For both primary outcomes we perform these tests at level $\alpha = 0.025$ using a Holm-Bonferroni correction for multiple comparisons.

One concern with this comparison, however, is that students who participate in high school sports may differ substantially from non-participants in terms of personality, temperament, and overall fitness and lifestyle (all of which are unmeasured). 
As such, including those subjects from the Non-Sport control group may introduce problematic unmeasured confounding. 
The Other Sport control group is arguably a closer and more appropriate control group.
Simply dropping the non-sport controls from our analysis would cut our effective sample size by about 1/3 and could result in a substantial decrease in power. 
Moreover, additional comparisons between treated group and the Non-Sport controls and between both control groups could show that it is playing football, and not simply playing high school sports, that are driving the outcomes. 

That is, by using both control groups separately we may systematically vary the unmeasured confounders of concern.
In order to preserve the increased power of using controls from both groups while still testing the treated group against each group separately we consider an ordered testing procedure which controls the family wise error rate (FWER) \citep{Rosenbaum2008}. 
In particular, we first test the null of no treatment effect using matched sets constructed with all controls.
If we reject that at level alpha we conduct the same test separately using matched sets constructed using non-contact sport playing controls and controls who did not play sports. 
If we reject both separate tests at level alpha then we perform an equivalence test between the two control groups. 
If at any stage of the ordered testing procedure we do not reject, we stop the procedure. 
For example, if we do not reject the test using all controls we do not continue on to test against the two control groups separately. 
This stopping rule is what guarantees FWER control.

We will perform the above ordered testing procedure for both the composite cognitive outcome and the depression outcome. 
We will also report marginal 97.5\% confidence intervals that go along with each test (regardless of whether we reach the test in the testing in order procedure).
Table~\ref{tab:example_results} illustrates how we will report the results from this testing procedure. 
For the comparison of the two control groups, the alternative hypothesis is that the two groups differ less from each other than either group differs from the treated group.
Note, the entires in Table~\ref{tab:example_results} are for illustrative purposes only. 
\begin{table}[H]
\centering
\caption{Estimated treatment effect and marginal 97.5\% confidence interval for each comparison. $* = $ significant using the testing in order procedure that controls FWER at 0.05.}
\label{tab:example_results}
\small
\begin{tabular}{lcccc}
\hline
\multirow{2}{*}{Outcome}
& Football vs & Football vs & Football vs & Non-Sport Controls vs \\
~ & All Controls & Non-Sport Controls & Other Sport Controls & Other Sport Controls \\ \hline
Cognitive Functioning & -0.6 (-1.5, 0.3) & -0.3 (-2.0, 1.7) & -1.3 (-2.5, -0.1) & -.08 (-2.0, 0.4) \\
Depression & 1.4 (0.3, 1.8)* & 0.9  (-0.6, 2.4) & 1.8 (0.2, 3.4)* & 0.6 (-0.8, 1.8) \\ \hline
\end{tabular}

\end{table}

To carry out this testing in order procedure, we build four separate matches
\begin{itemize}
\item{Match 1: Football players to all controls}
\item{Match 2: Football players to Non-Sport controls}
\item{Match 3: Football players to Other Sport controls}
\item{Match 4: Non-Sport controls to Other Sport controls}

\end{itemize}

For Match 1, each matched set consists of exactly one football player and up to 6 controls.
For Matches 2 and 3, each matched set consists of either one football players and up to 6 controls or one control and up to three football players.
For Match 4, each matched set consisted of one subject from one control group and up to six from the other.
The tables in Appendix~\ref{sec:matched_set_structure} list the composition of the matched sets for these four matches.
To assess whether each match produced adequate covariance balance, we compute standardized differences, which are shown in the Love plots in Appendix~\ref{sec:matched_set_structure}.

Looking at the Love plot from Match 1, we see that prior to matching, football players tended to come from smaller schools (negative standardized difference for the covariate \textbf{hssize}) and were more likely to be involved in student government (positive standardized difference for the covariate \textbf{schgovt}). After matching, however, we achieve good balance on these two covariates. We see a similar pattern with Match 2. In Match 3, however, we note that the covariates were already balanced before matching. It is reassuring, then, to see that the covariates remained in balance after matching. In Match 4, we observe that there is substantial imbalance for \textbf{hssize} and \textbf{schgovt} prior to matching. 

\section{Secondary Analysis}
\label{sec:secondary_analysis}

Secondary and intermediate outcomes will be analyzed using the matched sets from Match 1 (i.e. we will not use a testing in order procedure).  
We will report the marginal p-value and also whether the p-value is significant after adjusting for multiplicity with the Benjamini-Hochberg procedure.  
For continuous outcomes, we will use the same method as for the primary outcomes of regressing the outcome on football status, matched set dummies and the covariates.  For the binary secondary outcomes, we will use conditional logistic regression of the outcome on football status and the covariates with the matched sets being strata. 

Football playing status is an admittedly coarse measure of potential exposure to repetitive head trauma. A much finer analysis \footnote{Finer still would be to use the number of practices or games played but this data is not recorded} would account for the length of exposure. 
To estimate this dose effect on our primary outcomes, we will repeat our primary analysis but use the number of years of football participation (as recorded by a studentÕs yearbook) as a dose. 
Functionally, this involves re-running our primary analysis but scaling the response variables by the number of years of football participation.

\section{Sensitivity Analysis}
\label{sec:sensitivity_analysis}

Controlling for observed baseline covariates through matching is designed to eliminate bias in treatment assignment by balancing the distribution of observable potential confounders between the treated and control groups. 
However, it cannot ensure balance of unobserved confounders unless they are highly correlated with observed confounders. 
Conditional on the full matching, let $\Gamma$ bound the odds ratio of treatment for any pair in a matched set. 
For each marginally significant outcome in both the primary and secondary analysis we will report the $\Gamma$ at which the result is sensitive (i.e. the $\Gamma$ at which the result becomes insignificant). 
This is known as a sensitivity analysis \citep{Rosenbaum2002} and allows us to quantitatively assess how sensitive the results are to bias in the treatment assignment due to the observational nature of the study.  Larger values of $\Gamma$ provide greater evidence for the studyÕs causal conclusions. 
Sensitivity analyses for the continuous primary and secondary outcomes will be performed with the \texttt{sensitivitymv} package in \texttt{R} where we will use the Huber-Maritz M-test with no trimming at level $\alpha =.025$ for the primary outcomes and $\alpha=.05$ for the secondary outcomes.  
The sensitivity analysis will be performed on the residuals after regressing the outcomes on the covariates -- this is the covariance adjustment procedure suggested by \citep{Rosenbaum2002a}. 
For the binary secondary outcomes, we will use a sensitivity analysis for testing the null hypothesis of no treatment effect using the Mantel-Haenszel test \citep[Section 4.2]{Rosenbaum2002}.

\section{Acknowledgements}
\label{sec:acknowledgements}

We thank Paul Rosenbaum for helpful comments and suggestions on our work.

\newpage
\bibliography{football_cognitive_depression}

\newpage
\appendix
\section{Baseline Covariates}
\label{sec:baseline_covariates}
We use the following baseline covariates to match subjects. In parentheses, we list the corresponding variable name in the WLS dataset. Covariates marked with a $^\dagger$ were constructed from the WLS variable listed. For instance, the variable \textbf{schgovt} records the number of student government-related activities listed in each subjectÕs senior yearbook. We convert these counts to a binary variable indicating whether or not the student participated in at least one student government activity. The WLS variable \textbf{plns58q} records responses to the survey question ÒWhat are your plans for next year?Ó All covariates were collected in 1957, except for those marked with $^\ddagger$, which were recorded in 1975.

\begin{itemize}
\item{Extent to which subject discussed plans with teachers or counselors (\textbf{tchncntq})}
\item{Extent to which subject discussed plans with parents (\textbf{parcntq})}
\item{Whether teachers encouraged subject to attend collect (\textbf{tcheneq})}
\item{Whether parents wanted subject to attend college (\textbf{parencq})}
\item{1970 Duncan SEI score for the job to which subject aspired in 1957 (\textbf{sposcasp3})}
\item{How subject's family income or wealth compared to that of families in his community (\textbf{sesp57})}
\item{1957 high school class size (\textbf{assize})}
\item{Whether teachers considered subject an outstanding student (\textbf{tchevl})}
\item{Whether parents are able to help subject go to college financially (\textbf{parsup})}
\item{Parent's income in 1957 (\textbf{bmpin1})}
\item{Whether subject participated in orchestra, band, chorus, or other smaller musical ensembles (\textbf{musperf})$^\dagger$}
\item{Whether subject participated in any high school drama, speech, or debate activities (\textbf{spchperf})$^\dagger$}
\item{Whether subject participated in any student government activities (\textbf{schgovt})$^\dagger$}
\item{Whether subject was involved in any school publications (\textbf{schpubs})$^\dagger$}
\item{1957 population of town in which subject attended high school, with rural-urban distinction (\textbf{rlur57}).}
\item{Whether subject planned to join military (\textbf{plns58q})$^\dagger$}
\item{Whether subject attended a Catholic high school (\textbf{hsmd57})$^\dagger$}
\item{IQ scored mapped from raw Henmon-Nelson scores (\textbf{gwiiq\_bm})}
\item{Parent's education (\textbf{bmfaedu, bmmaedu})$^\ddagger$}
\item{Whether subject lived with both parents most of the time up until 1957 (\textbf{bklvpr})$\ddagger$}
\item{Whether mother had a job in 1957 (\textbf{wrmo57})$^\ddagger$}
\item{Number of years of further education subject had planned to get in 1957 (\textbf{zpedyr})$^\ddagger$}
\item{Whether subject's friends planned to go to college in 1957 (\textbf{zfrplc})$^\ddagger$}

\end{itemize}

\section{Composition of Matched Sets}
\label{sec:matched_set_structure}

For Matches 1,2, and 3, the first number in the composition column in the tables below is the number of treated subjects in a matched set and the second number gives the number of control subjects in the matched set.
For Match 4, the first number in the composition column is the number of Non-Sport controls in the matched set and the second is the number of Other Sport controls in the matched set.

\begin{table}[H]
\centering
\caption{The number of matched sets from Match 1 of each composition}
\label{tab:match1_composition}
\begin{tabular}{lccccccc} \hline
Composition & LF, DWR, CES-D & LF,DWR & LF, CES-D & DWR, CES-D & LF & DWR & CES-D \\ \hline
3:1 & 0 & 0 & 0 & 0 & 0 & 0 & 0 \\
2:1 & 0 & 0 & 0 & 0 & 0 & 0 & 0 \\
1:1 & 324 & 32 & 14 & 31 & 6 & 8 & 135 \\
1:2 & 23 & 3 & 3 & 9 & 0 & 2 & 12 \\
1:3 & 24 & 0 & 2 & 4 & 1 & 1 & 9 \\
1:4 & 12 & 2 & 0 & 1 & 0 & 1 & 9 \\
1:5 & 11 & 2 & 0 & 0 & 0 & 1 & 1 \\
1:6 & 73 & 17 & 5 & 10 & 2 & 0 & 44 \\ \hline
\end{tabular}
\end{table}

\begin{table}[H]
\centering
\caption{The number of matched sets from Match 2 of each composition}
\label{tab:match2_composition}
\begin{tabular}{lccccccc} \hline
Composition & LF, DWR, CES-D & LF,DWR & LF, CES-D & DWR, CES-D & LF & DWR & CES-D \\ \hline
3:1 & 60 & 7 & 1 & 5 & 0 & 1 & 23 \\
2:1 & 5 & 0 & 2 & 2 & 2 & 2 & 4 \\
1:1 & 182 & 15 & 12 & 21 & 3 & 3 & 79 \\
1:2 & 21 & 5 & 2 & 2 & 0 & 2 & 17 \\
1:3 & 7 & 1 & 0 & 3 & 0 & 0 & 5 \\
1:4 & 11 & 0 & 0 & 3 & 0 & 1 & 5 \\
1:5 & 8 & 1 & 0 & 3 & 0 & 0 & 5 \\
1:6 & 73 & 17 & 5 & 10 & 2 & 0 & 22 \\ \hline
\end{tabular}
\end{table}

\begin{table}[H]
\centering
\caption{The number of matched sets from Match 3 of each composition}
\label{tab:match3_composition}
\begin{tabular}{lccccccc} \hline
Composition & LF, DWR, CES-D & LF,DWR & LF, CES-D & DWR, CES-D & LF & DWR & CES-D \\ \hline
3:1 & 73 & 10 & 5 & 0 & 2 & 0 & 25 \\
2:1 & 33 & 3 & 1 & 1 & 1 & 3 & 16 \\
1:1 & 175 & 17 & 5 & 18 & 1 & 7 & 95 \\
1:2 & 3 & 1 & 1 & 3 & 0 & 0 & 4 \\
1:3 & 2 & 1 & 0 & 10 & 0 & 0 & 2 \\
1:4 & 2 & 0 & 0 & 0 & 0 & 0 & 1 \\
1:5 & 0 & 1 & 0 & 0 & 0 & 0 & 1 \\
1:6 & 0 & 0 & 1 & 0 & 0 & 0 & 0 \\ \hline
\end{tabular}
\end{table}

\begin{table}[H]
\centering
\caption{The number of matched sets from Match 4 of each composition}
\label{tab:match4_composition}
\begin{tabular}{lccccccc} \hline
Composition & LF, DWR, CES-D & LF,DWR & LF, CES-D & DWR, CES-D & LF & DWR & CES-D \\ \hline
6:1 & 1 & 0 & 0 & 0 & 0 & 0 & 0 \\
5:1 & 1 & 0 & 1 & 0 & 0 & 0 & 0 \\
4:1 & 4 & 0 & 0 & 1 & 0 & 0 & 1 \\
3:1 & 4 & 1 & 0 & 0 & 0 & 0 & 1 \\
2:1 & 10 & 2 & 0 & 1 & 0 & 2 & 5 \\
1:1 & 133 & 11 & 5 & 14 & 0 & 4 & 92 \\
1:2 & 17 & 2 & 4 & 0 & 1 & 1 & 8 \\
1:3 & 8 & 3 & 0 & 0 & 1 & 0 & 7 \\
1:4 & 13 & 5 & 3 & 1 & 0 & 0 & 5 \\
1:5 & 7 & 1 & 1 & 0 & 0 & 0 & 4 \\
1:6 & 64 & 11 & 1 & 12 & 2 & 1 & 26 \\ \hline
\end{tabular}
\end{table}

\begin{figure}[H]
\centering
\begin{subfigure}[b]{0.48\textwidth}
\centering
\includegraphics[width = \textwidth]{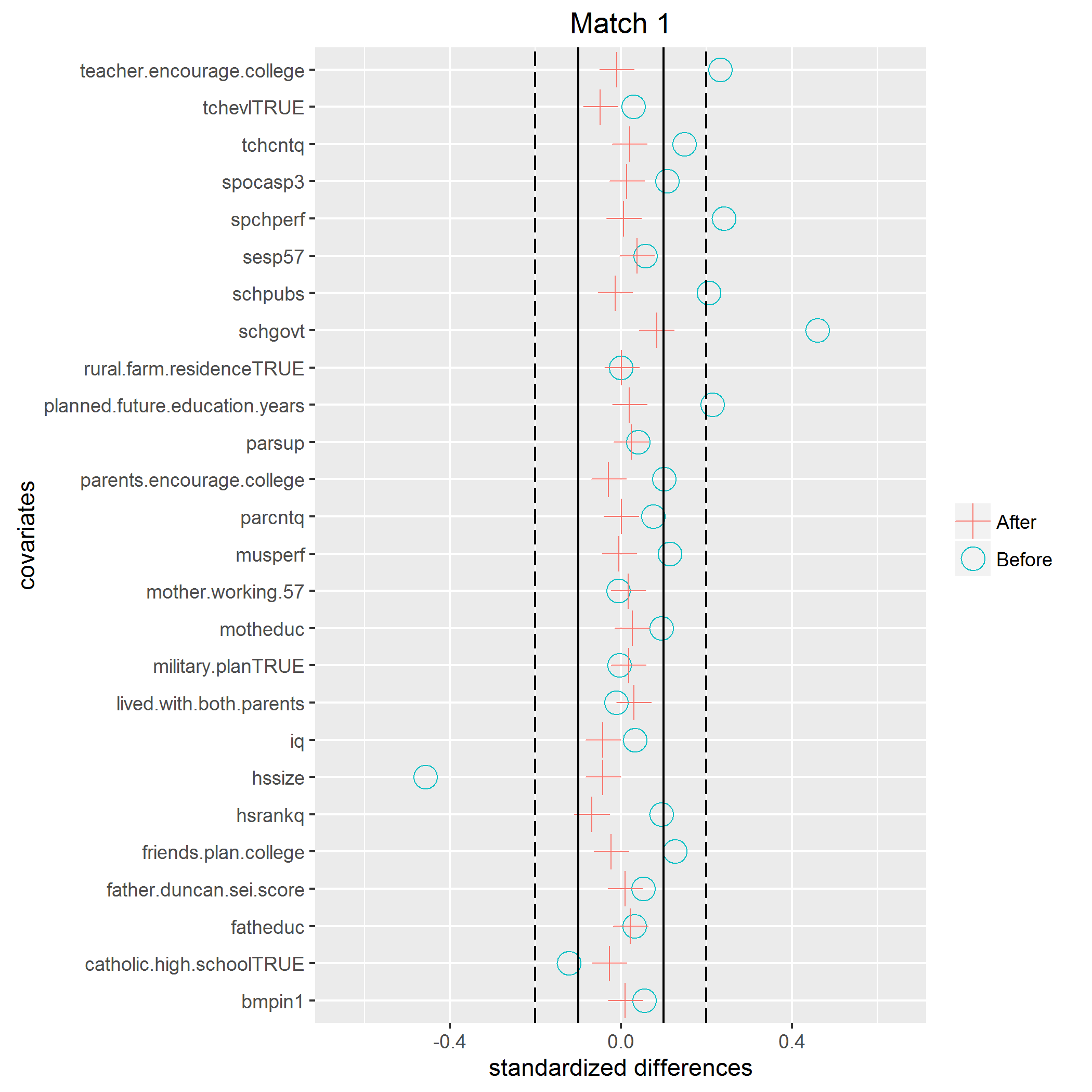}
\end{subfigure}
\begin{subfigure}[b]{0.48\textwidth}
\includegraphics[width = \textwidth]{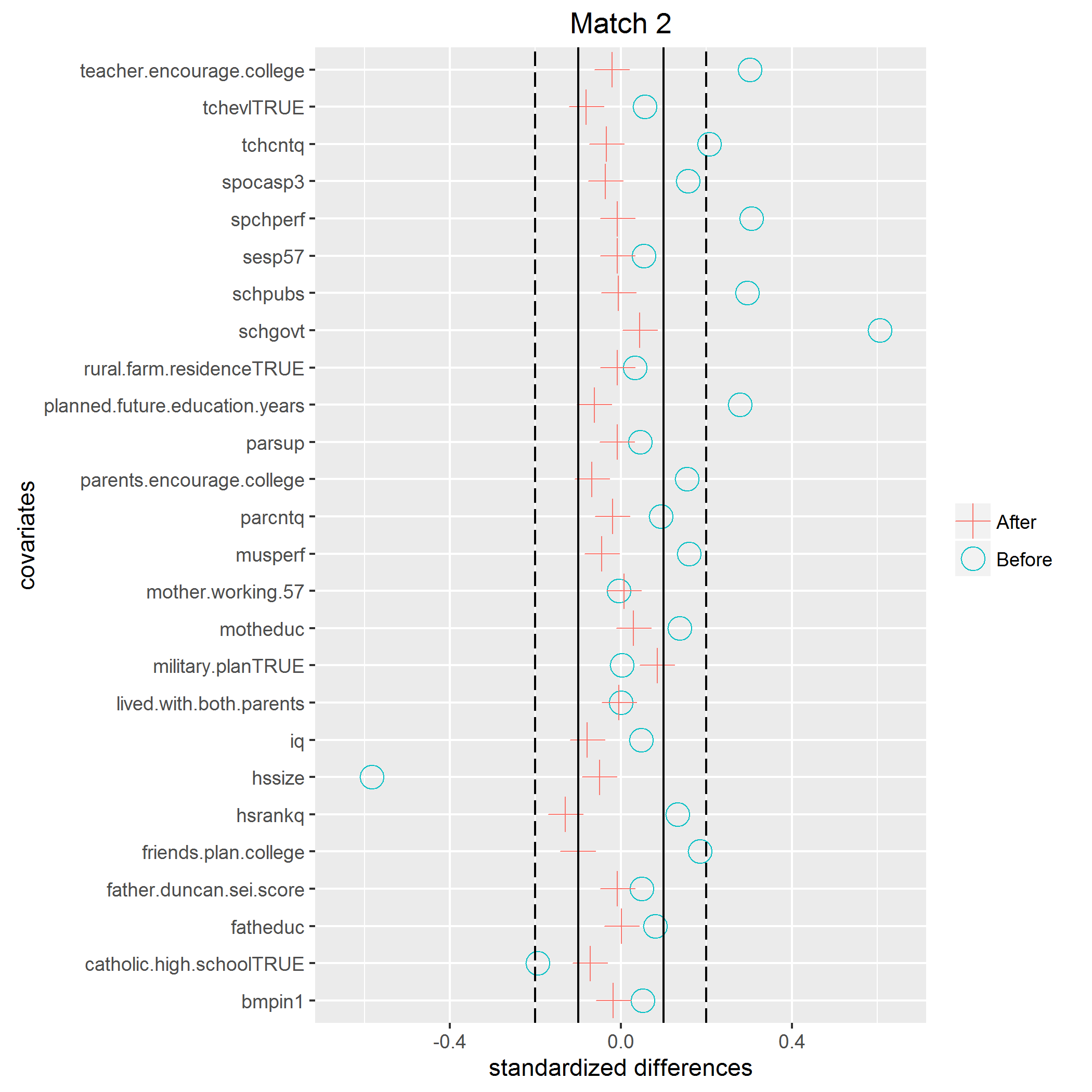}
\end{subfigure}
\\
\begin{subfigure}[b]{0.48\textwidth}
\includegraphics[width = \textwidth]{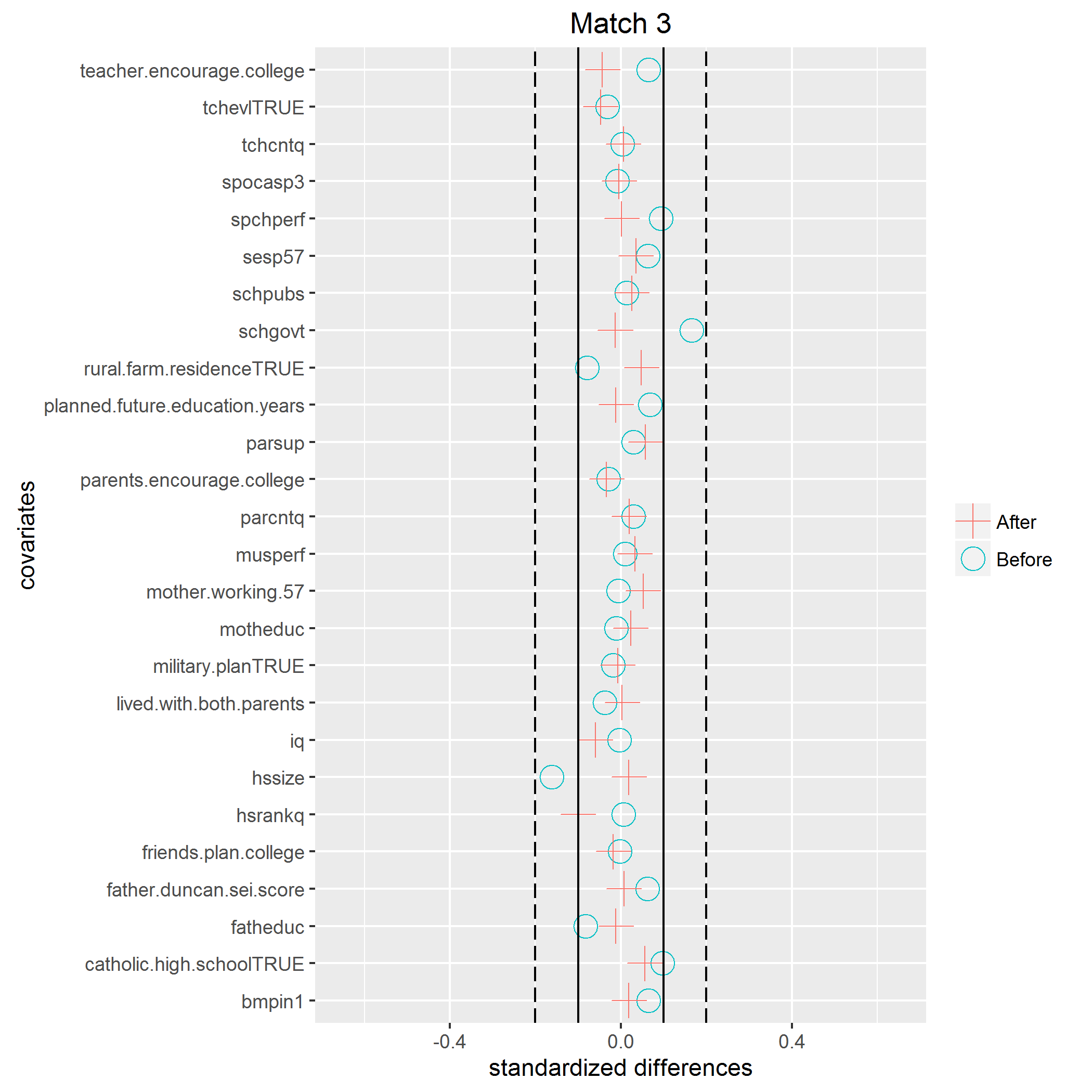}
\end{subfigure}
\begin{subfigure}[b]{0.48\textwidth}
\includegraphics[width = \textwidth]{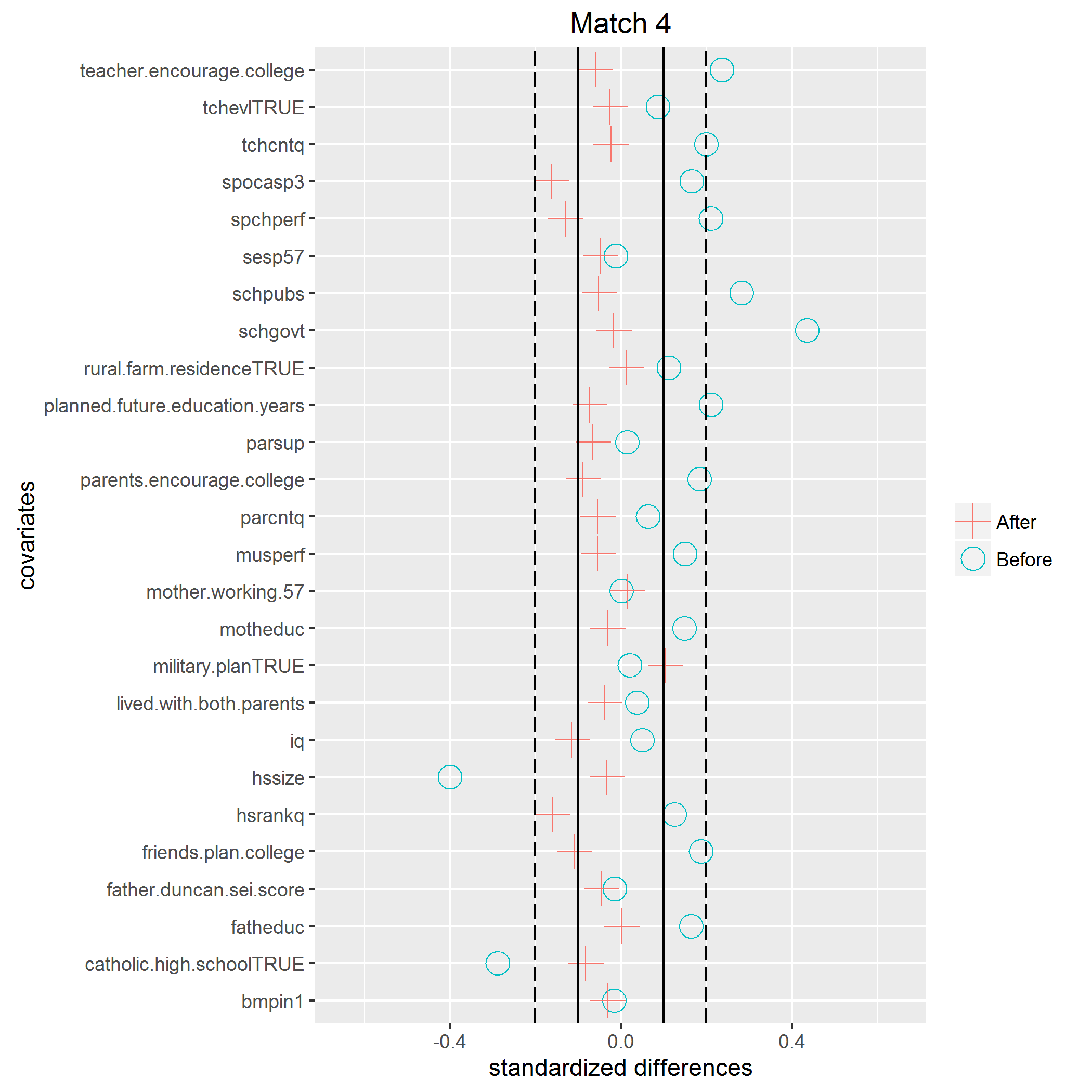}
\end{subfigure}
\end{figure}

\end{document}